\documentstyle[12pt,aaspp4]{article}

\newcommand{\be}{\begin{equation}}
\newcommand{\ee}{\end{equation}}
\newcommand{\lab}[1]{\label{#1}}
\newcommand{\rr}[1]{~(\ref{#1})}

\begin{document}
\title{Harmonizing the RR Lyrae and Clump Distance Scales --- Stretching the
Short Distance Scale to Intermediate Ranges?}
\author{Piotr Popowski}
\affil{Institute of Geophysics and Planetary Physics, L-413\\ Lawrence Livermore
National Laboratory, University of California\\ P.O.Box 808, Livermore, CA 94551, USA.\\ E-mail: popowski@igpp.llnl.gov.}

\begin{abstract}
I explore the consequences of making the RR Lyrae and clump
giant distance scales consistent in the solar neighborhood, Galactic bulge
and Large Magellanic Cloud (LMC). I employ two major assumptions:
1) that the absolute magnitude - metallicity, $M_V(RR)$ -- [Fe/H], relation for
RR Lyrae stars is universal, and 2) that absolute $I$-magnitudes of clump
giants, $M_I(RC)$, in Baade's Window are known (e.g., can be inferred from the 
local Hipparcos-based calibration or theoretical modeling). A comparison 
between the solar neighborhood and Baade's Window sets $M_V(RR)$ at 
[Fe/H] = $-1.6$ in the range $(0.59 \pm 0.05, 0.70 \pm 0.05)$, somewhat 
brighter than the statistical parallax solution. More luminous RR Lyrae stars 
imply younger ages of globular cluster, which would be in better agreement 
with the conclusions from the currently favored stellar evolution and 
cosmological models.
A comparison between Baade's Window and the LMC sets the $M_I^{\rm LMC}(RC)$ 
in the range $(-0.33 \pm 0.09, -0.53 \pm 0.09)$. The distance modulus to 
the LMC is $\mu^{\rm LMC} \in (18.24 \pm 0.08, 18.44 \pm 0.07)$. 
Unlike $M_I^{\rm LMC}(RC)$, this range in $\mu^{\rm LMC}$ does {\em not} depend
on the adopted value of the dereddened LMC clump magnitude, 
$I_0^{\rm LMC}(RC)$.
I argue that the currently available information is insufficient to select 
the correct distance scale with high confidence.

Subject Headings: distance scale --- dust, extinction --- Galaxy: center --- Magellanic Clouds --- stars: horizontal-branch --- stars: variables: RR Lyrae

\end{abstract}

\section{Introduction}

The Hubble Space Telescope Key Project (e.g., Madore et al.\ 1999) 
concluded that the biggest uncertainty in the Hubble constant, $H_0$,
comes from the uncertainty in the distance to the LMC. 
Among the major methods that have been used to 
determine the distance to the LMC are:
the echo of the supernova 1987A, solving parameters of eclipsing
binaries, Cepheids, RR Lyrae stars, and red clump giants.
They all suffer from some uncertainties and possible systematic errors.
The echo of the supernova 1987A was a transient event with limited data and 
contradictory interpretations (Gould \& Uza 1998 versus Panagia 1998). 
Only one attempt of solving eclipsing binary using space-based spectra was made
by Guinan et al.\ (1998) for HV 2274. Their result is sensitive to the 
reddening toward HV 2274 (Udalski et al.\ 1998 versus Nelson et al.\ 2000).
To be calibrated with high precision, Cepheids have to wait for the next 
generation astrometric missions (for the Hipparcos-based calibration see
Feast \& Catchpole 1997 and Pont 1999).
The absolute $V$-magnitudes of RR Lyrae stars, $M_V(RR)$, are still
under debate with a faint value given by the statistical parallax method
and a bright value suggested by the main sequence fitting 
(see Popowski \& Gould 1999). 
The major problem of the red clump method is the possibility
that the absolute $I$-magnitude, $M_I(RC)$, is sensitive to the environment
(Cole 1998; Girardi et al.\ 1998; Twarog, Anthony-Twarog, \& Bricker 1999).
The mentioned methods give results inconsistent within their estimated
uncertainties, which suggests hidden systematics.

Here I concentrate on two horizontal-branch standard candles:
red clump giants and RR Lyrae stars. I start with a very short review
of their application to determine the distance to the LMC.
Paczy\'{n}ski \& Stanek (1998) pointed out that clump giants
should constitute an accurate distance indicator. 
In a study of the morphology of the red clump, 
Beaulieu \& Sackett (1998) argued that a distance modulus of 
$\mu^{\rm LMC}=18.3$
provides the best fit to the dereddened LMC color-magnitude diagram. 
Udalski et al.\ (1998a) and 
Stanek, Zaritsky, \& Harris (1998) applied the I-magnitude based approach
of Paczy\'{n}ski and Stanek (1998) and found a very 
short distance to the LMC ($\mu^{LMC} \approx 18.1$). In response, Cole 
(1998) and Girardi et al.\ (1998) suggested that clump giants are not standard 
candles and that their $M_I(RC)$ depend on the 
metallicity and age of the population. Udalski (1998b, 1998c) countered this 
criticism by showing that the metallicity dependence is at a low level of 
about $0.1$ mag/dex, and that the $M_I(RC)$ is approximately constant for 
cluster ages between 2 and 10 Gyr. The new determinations of the $M_I(RC)$
-- [Fe/H] relation by Stanek et al.\ (2000), Udalski (2000) and 
Popowski (2000) indicate 
a moderate slope of $0.10-0.20$ mag/dex.
The only clump determination, which results in a truly long distance
to the LMC is a study by Romaniello et al.\ (2000) who investigated the field 
around supernova SN 1987A, which is not well suited for extinction
determinations.
Romaniello et al.\ (2000) also assumed a bright $M_I(RC)$ from 
theoretical models.
To address the issue of possible extinction overestimate in earlier studies
(see e.g., Zaritsky 1999 for a discussion), Udalski (1998c, 2000) measured
clump magnitudes in low extinction regions in and around the LMC clusters. 
The resulting $\mu^{LMC} = 18.24 \pm 0.08$ (Udalski 2000) is often
perceived as the least model-dependent distance modulus to the LMC obtained
from clump giants.

Different methods to determine the RR Lyrae absolute magnitude are analyzed 
in Popowski \& Gould (1999). The results depend on the methods used.
When the kinematic or geometric determinations are employed,
one obtains $M_V(RR) = 0.71 \pm 0.07$ at [Fe/H] $= -1.6$ (with 
$M_V(RR) = 0.77 \pm 0.13$ from the best understood method, statistical 
parallax).
The other methods typically produce or are consistent with brighter
values. The representative main sequence fitting to globular clusters
gives $M_V(RR) = 0.45 \pm 0.12$ at [Fe/H] $= -1.6$ (Carretta et al.\ 2000).
When coupled with Walker (1992) photometry of globular clusters, Popowski
\& Gould's (1999) best $M_V(RR)$  results in $\mu^{LMC} = 18.33 \pm 0.08$.
When Udalski et al.\ (1999) photometry of the LMC field RR Lyrae 
stars is used, one obtains $\mu^{LMC}= 18.23 \pm 0.08$.

The essence of the approach presented here is a comparison
between clump giants and RR Lyrae stars in different environments.
If answers from two distance indicators agree then either
the systematics have been reduced to negligible levels in both of them
or the biases conspire to produce the same answer.
This last problem can be tested with an attempt to synchronize distance
scales in three different environments, because a conspiracy of systematic 
errors is not likely to repeat in all environments.
Here I show that combining the information on RR Lyrae and red clump stars
in the solar neighborhood, Galactic bulge, and LMC provides additional
constraints on the local distance scale.

\section{Assumptions and Observational Data}

The results I present in \S 3 and \S 4 are not entirely general and have been 
obtained
based on certain theoretical assumptions about the nature of standard candles
and populations in different stellar systems. In addition, the conclusions
depend on the source of photometry. One does not have much freedom in this 
regard, but I have made certain choices, which I describe in \S 2.2.

\subsection{Theoretical assumptions}

This investigation relies strongly on the following two
assumptions:
\begin{enumerate}
\item The $M_V(RR)$ -- [Fe/H] relation for RR Lyrae stars is universal.
More specifically, I assume that for every considered system, $M_V(RR)$ is 
only a linear function
of this system's metallicity:
\be
M_V(RR) = \alpha \left( {\rm [Fe/H]} + 1.6 \right) + \beta . \lab{mvfeh}
\ee
Moreover, I will assume that the slope  $\alpha = 0.18 \pm 0.03$, which is not 
critical for the method but determines the numerical results.
In the most general case, $M_V(RR)$ depends on morphology of the horizontal 
branch (Lee, Demarque, \& Zinn 1990; Caputo et al.\ 1993). However, for average
non-extreme environments (here the character of environment can be judged
using the Lee 1989 index) a linear, universal $M_V(RR)$ -- [Fe/H] should be a 
reasonable description. For the RR Lyrae stars of the Galactic halo (either in
the solar neighborhood or in Baade's Window) and of the LMC field or globular 
clusters, equation\rr{mvfeh} with universal $\alpha$ and $\beta$ should 
approximately hold.
The universal character of the calibration is essential to any distance
determination with standard candles, and so this assumption is rather
standard.
\item The absolute magnitude $M_I^{\rm BW}(RC)$ of the bulge clump giants 
is known, which in practice means one of two things: either one takes
the results of population modeling or infers the value from the 
Hipparcos-calibrated $M_I^{\rm HIP}(RC)$ of the local clump stars.
I will temporarily adopt the second route and assume that there are 
no population factors except metallicity that influence $M_I^{\rm BW}(RC)$
in the Galactic bulge (with respect to the local clump) or that their 
contributions cancel out.
Again, this is somewhat similar to point 1., but here I am more flexible 
allowing $M_I^{\rm LMC}(RC)$ in the LMC not to follow the local Hipparcos 
calibration (that is, I allow population effects of all types).
\end{enumerate}

\subsection{Data}

The calibration of clump giants in the solar neighborhood is based on Hipparcos
(Perryman 1997) data for nearly 300 clump giants as reported by 
Stanek \& Garnavich (1998) and refined by Udalski (2000).
\be
M_I^{\rm HIP}(RC) = (-0.26 \pm 0.02) + (0.13 \pm 0.07) ({\rm [Fe/H]}+0.25) \lab{mihip}
\ee
I assume that the metallicity of the bulge clump in Baade's Window
is [Fe/H] $= 0.0 \pm 0.3$ consistent with Minniti et al.\ (1995). 
As a result, I set $M_I^{\rm BW}(RC) = -0.23 \pm 0.04$ (see eq.\rr{mihip} 
and \S 2.1), where the error of $0.04$ is dominated by the uncertainty
in the metallicity of clump giants in Baade's Window. I stress that one
can simply assume $M_I^{\rm BW}(RC)$ without any reference to Hipparcos results
and obtain the conclusions reported later in Table 1. Equation (2) and the 
following considerations serve only as the evidence that, in the lack of 
significant population effects, this choice of $M_I^{\rm BW}(RC)$ would be 
well justified.

The $V$- and $I$-band photometry for the bulge clump giants and RR Lyrae stars
originates from, or have been calibrated to the photometric zero-points of,
phase-II of the Optical Gravitational Lensing Experiment (OGLE). That is, the data for Baade's
Window come from Udalski (1998b) and were adjusted according to
zero-point corrections given by Paczy\'{n}ski et al.\ (1999).
When taken at face value, these data result in $(V-I)_0$ 
colors\footnote{Here and thereafter subscript ``0'' indicates dereddened or 
extinction-free value.} of both
clump giant and RR Lyrae stars that are 0.11 redder than for their local
counterparts. To further describe the input data let me define
$\Delta$ for a given stellar system as the difference between the 
mean dereddened I-magnitude of clump giants and the derredened V-magnitude of 
RR Lyrae stars at the metallicity of RR Lyrae stars in the Galactic bulge.
The quantity $\Delta$ allows one to compare the relative brightness of clump 
giants
and RR Lyrae stars in different environments and so will be very useful
for this study (for more discussion see Udalski 1998b and Popowski 2000).
In the Baade's Window with anomalous horizontal branch colors 
$\Delta^{\rm BW} \equiv I^{\rm BW}_0(RC)-V^{\rm BW}_0(RR) = -1.04 \pm 0.04$.
When the color correction considered by Popowski (2000) is taken into
account one obtains $\Delta^{\rm BW} = -0.93 \pm 0.04$.

In the LMC, I use dereddened $I_0 = 17.91 \pm 0.05$
for ``representative red clump''. Here ``representative'' means in clusters
(compare to $I_0 = 17.88 \pm 0.05$ from Udalski 1998c)
or in fields around clusters (compare to $I_0=17.94 \pm 0.05$ from 
Udalski 2000).
The advantage of using $I_0$ from cluster and cluster fields 
is their low, well-controlled extinction (Udalski 1998c, 2000).
I take $V_0=18.94 \pm 0.04$ for field RR Lyrae stars at 
[Fe/H] $= -1.6$ from 
Udalski et al.\ (1999) and adopt $V_0=18.98 \pm 0.03$ at [Fe/H] $= -1.9$ for 
the cluster RR Lyrae stars investigated by Walker (1992).
The difference of photometry between 
Udalski et al.\ (1999) and Walker (1992)
may have several sources. The least likely is that the cluster system is
displaced with respect to the center of mass of the LMC field. 
Also, cluster RR Lyrae stars could be intrinsically 
fainter, but again this is not very probable. 
I conclude that the difference comes either from 1) extinction, or
2) the zero-points of photometry.
The first case would probably point to overestimation of extinction by OGLE, 
because it is harder to determine the exact extinction in the field than 
it is in the clusters. The second case can be tested with independent
LMC photometry.
In any case, the difference of $\sim 0.1$ mag is an indication of 
how well we currently measure $V_0(RR)$ in the LMC.

Finally, let us note that the homogeneity of photometric data was absolutely 
essential for the investigation of the global slope in the $M_I(RC)$ -- [Fe/H] 
relation (Popowski 2000). Here it is not as critical. Still, the common source
of data for the Galactic bulge reduces the uncertainty in the $M_V(RR)$ 
calibration. On the other hand, the use of both OGLE and Walker's (1992) data 
for the LMC quantifies a possible level of extinction/photometry uncertainty.

\section{The method and results}

The distance modulus to the Galactic center from RR Lyrae stars is:
\be
\mu^{\rm BW}(RR) = V^{\rm BW}_0(RR) - M_V^{\rm BW}(RR).
\lab{rrdist}
\ee
I assume the RR Lyrae metallicities of ${\rm [Fe/H]}_{RR}^{\rm BW} = -1.0$ 
from Walker \& Terndrup (1991).
The distance modulus to the Galactic center from the red clump can be 
expressed as:
\be
\mu^{\rm BW}(RC) = I^{\rm BW}_0(RC) - M_I^{\rm BW}(RC).
\lab{clumpdist}
\ee
The condition that $\mu^{\rm BW}(RR)$ and $\mu^{\rm BW}(RC)$ are equal to each 
other\footnote{For this condition to be exactly true one has to take
into account the distribution of clump giants in the bar and RR Lyrae
stars in the spheroidal system as well as completeness characteristics 
of a survey. The analyses from OGLE did not reach this level of detail,
but I neglect this small correction here.} results in:
\be
M_I^{\rm BW}(RC)-M_V^{\rm BW}(RR)=I^{\rm BW}_0(RC)-V^{\rm BW}_0(RR)
\label{differences}
\ee
But the right hand side of equation\rr{differences} is just $\Delta^{BW}$,
which is either directly
taken from dereddened data or determined by solving the color
problem (for more detail see Popowski 2000).
If there are no
population differences between the clump in Baade's Window and the solar
neighborhood (as we assumed in \S 2.1), then $M_I^{\rm BW}(RC)$ is extremely 
well constrained from the Hipparcos results reported in equation\rr{mihip}.
Therefore, equation\rr{differences} is in effect the calibration of the 
absolute magnitude of RR Lyrae stars:
\be
M_V^{\rm BW}(RR) = M_I^{\rm BW}(RC) - \Delta^{BW}
\label{mvrr}
\ee
If one calibrates the $M_V(RR)$ -- [Fe/H] relations according to 
equation\rr{mvrr}, then by construction the solar neighborhood's and the 
Baade's Window's distance scales are consistent.

To determine $M_I^{\rm LMC}(RC)$, I construct the Udalski's (1998b) diagram.
However, both Udalski (1998b) and Popowski (2000) used such diagrams to
determine a global slope of the $M_I(RC)$ -- [Fe/H] relation. Because I am 
interested here just in the LMC, a more
powerful approach is to treat the Udalski (1998b) diagram in a discrete way.
That is, instead of fitting a line to a few points one takes a difference 
between the Baade's Window and LMC $\Delta$
as a measure of the $M_I(RC)$ differences in these two stellar systems.
Therefore:
\be
M^{\rm LMC}_I(RC) = M^{\rm BW}_I(RC) - (\Delta^{\rm BW} - \Delta^{\rm LMC})
\label{milmc}
\ee
The interesting feature of equation\rr{milmc} is that the calibration
of $M^{\rm LMC}_I(RC)$, even though based on RR Lyrae stars, is independent 
of the zero-point $\beta$ of the $M_V(RR)$ -- [Fe/H] relation.
Because $M^{\rm LMC}_I(RC)$ leads to a specific value of $\mu^{LMC}$, coupling
$\mu^{LMC}$ with the LMC RR Lyrae photometry allows one to calibrate the 
zero-point of the $M_V(RR)$ -- [Fe/H] relation. However this calibration is 
not independent of the one presented in equation\rr{mvrr} and so does not 
provide any additional information.

Using equations\rr{mvrr} and\rr{milmc}, I calibrate the zero point $\beta$ of 
$M_V(RR)$ -- [Fe/H] relation as well as $M_I^{\rm LMC}(RC)$ of clump giants 
in the LMC.
The solutions are listed in Table 1. Different assumptions about the color 
anomaly in the Galactic bulge and the use of either OGLE-II or Walker's (1992)
photometry in the LMC result in four classes of $[M_V(RR),M_I^{LMC}(RC)]$ 
solutions (column 1). Following argument from \S 2.2, I use one universal
$I_0$ for clump giants in 
the LMC (column 2). The brighter RR Lyrae photometry in the LMC comes from 
OGLE (Udalski et al.\ 1999) and the fainter from Walker (1992) [column 3]. 
In column 4, I report $\Delta^{\rm LMC}$, which has been inferred from columns
2 and 3 assuming the the slope $\alpha$ in the $M_V(RR)$ -- [Fe/H] relation
is 0.18. In column 5, I give $\Delta^{BW}$. The resulting
$M_V(RR)$ at [Fe/H] = $-1.6$, $M_I^{\rm LMC}(RC)$, and the LMC 
distance modulus are shown in columns 6, 7, and 8, respectively.

The sensitivity of the results to the theoretical assumptions from
\S 2 can summarized in the following equation:
\be
\delta \beta = \delta M_I^{\rm LMC}(RC) = - \delta \mu^{\rm LMC} = -0.6 \; (\alpha_{\rm true}-0.18) + (M_{I,{\rm true}}^{\rm BW}(RC) + 0.23), 
\lab{correction1}
\ee
where the three $\delta$-type terms indicate potential corrections, 
$\alpha_{\rm true}$ 
is a real slope in RR Lyrae $M_V(RR)$ - [Fe/H] relation and 
$M_{I,{\rm true}}^{\rm BW}(RC)$ is a true absolute magnitude of clump giants 
in the Bulge. The multiplying factor of 0.6 in the first term is a difference
between the solar neighborhood and Baade's Window metallicity of RR Lyrae 
stars.
The distance scale could be made longer with either a larger (steeper) 
slope $\alpha_{\rm true}$ or a brighter $M_{I,{\rm true}}^{\rm BW}(RC)$ value.
Very few $M_V(RR)$ - [Fe/H] relation determinations argue for slopes steeper 
than 0.3, and clump giants in the Galactic bulge, which are old, are expected 
to be on average somewhat fainter than the ones in the solar neighborhood.
To give an example of application of equation\rr{correction1} let us assume
$\alpha_{\rm true}=0.3$ (e.g., Sandage 1993), and $M_I^{\rm BW}(RC)= -0.15$
(Girardi \& Salaris 2000; inferred from their $\Delta M_I^{RC}$ in Table 4
without any adjustment for a small [Fe/H] mismatch). The first term would 
result in a correction of
$-0.07$ mag and the second term would contribute 0.08 mag. In this case 
the two corrections would almost entirely cancel out resulting in both 
$\beta$ and $M_I^{\rm LMC}(RC)$ being 0.01 mag fainter and $\mu^{\rm LMC}$ 
being 0.01 mag smaller.
Even if one ignores the $M_{I,{\rm true}}^{\rm BW}(RC)$ - related correction,
it is hard to make absolute magnitudes of RR Lyrae and clump stars brighter 
by more than $0.07$ mag. Consequently, the distance moduli to the LMC 
reported in Table 1 are unlikely to increase by more than $0.07$ mag
as a result of adjustment to the theoretical assumptions from \S 2.

Another interesting question is the sensitivity of the results reported 
in Table 1 to the deredenned magnitudes adopted for the LMC.
These dependences are described by the following equations:
\be
\delta M_I^{\rm LMC}(RC) = \left( I_{0,{\rm true}}^{\rm LMC}(RC) - 17.91 \right) - (V_{0,{\rm true}}^{\rm LMC}(RR)- V_{0}^{\rm LMC}(RR)),
\lab{correction2}
\ee
\be
\delta \mu^{\rm LMC} = (V_{0,{\rm true}}^{\rm LMC}(RR)- V_{0}^{\rm LMC}(RR)),
\lab{correction3}
\ee
where $V_{0}^{\rm LMC}(RR)$ is either Udalski et al. (1999) or Walker (1992)
value described in \S 2.2.
In this treatment, the obtained distance modulus to the LMC does not depend 
on the dereddened I-magnitudes of clump giants! This is very fortunate 
because of the unresolved 
observational controversy [$I_0^{\rm LMC}(RC) \sim 17.9$ from 
Udalski (1998c, 2000) versus $I_0^{\rm LMC}(RC) \sim 18.1$ from 
Zaritsky (1999) or Romaniello et al.\ (1999)].
Note that keeping current $V_{0}^{\rm LMC}(RR)$ and adopting fainter 
$I_0^{\rm LMC}(RC)$ would result in rather faint values of $M_I^{\rm LMC}(RC)
\in (-0.13, -0.33)$, in potential disagreement with population models
(see Girardi \& Salaris 2000). This may suggest that either Udalski's 
(1998c, 2000) dereddened clump magnitudes are more accurate or that
dereddened $V$-magnitudes for RR Lyrae stars need revision.

\section{Discussion}

Using RR Lyrae stars and clump giants, I showed that the requirement of 
consistency between standard candles in different environments is a powerful 
tool in calibrating absolute magnitudes and obtaining distances.
If the anomalous character of $(V-I)_0$ in Baade's Window is real (i.e.,
not caused by problems with photometry or misestimate of the coefficient
of selective extinction), then the distance scale tends to be shorter.
In particular, $M_V(RR) = 0.70 \pm 0.05$ at [Fe/H] = $-1.6$, and the distance
modulus to the LMC spans the range from 
$\mu^{LMC} = 18.24 \pm 0.08$ to $18.33 \pm 0.07$.
If $(V-I)_0$ color of stars in Baade's Window is in error and should be
standard, then the distance scale is longer. In particular, one can
obtain $M_V(RR) = 
0.59 \pm 0.05$ at [Fe/H] = $-1.6$ and the distance modulus from 
$\mu^{LMC}= 18.35 \pm 0.08$ to $18.44 \pm 0.07$.
It is important to notice that the reported distance modulus ranges
do {\em not} change with the assumed value of the dereddened $I$-magnitudes
of the LMC clump giants, $I_0^{\rm LMC}(RC)$.

Are there any additional constraints that would allow one to select
the preferred value for RR Lyrae zero point $\beta$, $M_I^{\rm LMC}(RC)$, and 
$\mu^{\rm LMC}$?
The fact that indirectly favors the intermediate distance scale 
($\mu^{\rm LMC} \sim 18.4$) is 
its consistency with the results from classical Cepheids.
The value of $M_V(RR)$ required for such solution is only $1.4\,\sigma$ 
(combined) below 
the ``kinematic'' 
value of Popowski \& Gould (1999) and $1.3\,\sigma$ (combined) below the 
statistical parallax result given by Gould \& Popowski (1998),
leaving us without a decisive hint.
The Twarog et al.\ (1999) study of two open Galactic
clusters (NGC 2420 and NGC 2506) indicates rather bright red clumps.
However, the relevance of this result to the LMC is uncertain and, 
more importantly, its precision is too low to provide significant 
information.
The Beaulieu and Sackett (1998) study of clump morphology in the LMC suggests
$\mu^{LMC} \sim 18.3$, probably consistent with the entire (18.24, 18.44)
range.

The only significant but ambiguous clue is provided by Udalski's (2000) 
spectroscopically-based investigation of the red clump in the solar 
neighborhood.
One may entertain the following argument. 
If uncorrelated metallicity and age are the only population effects
influencing $M_I(RC)$ in different environments (with age argued to have
no effect in this case --- Udalski 1998c), then Hipparcos based calibration 
combined with $M_I^{\rm LMC}(RC)$ would naturally lead to an estimate of 
average metallicity of clump giants in the LMC. The brightest 
$M_I^{\rm LMC}(RC) = -0.53$
from Table 1 would result in ${\rm [Fe/H]}^{\rm LMC}=  -2.33$! Such a low
value is in violent disagreement with observations.
Therefore, either uncorrelated metallicity and age are not the 
only population effects influencing $M_I(RC)$ (see Girardi \& Salaris 2000 
for a discussion) or Udalski (2000) results coupled with typical LMC 
metallicities lend strong support to the shorter distance scale.
However, unless the selective extinction coefficient toward 
Baade's Window is unusual, very short distance scale comes at a price of 
anomalous $(V-I)_0$ bulge colors.
Therefore, one is tempted to ask: ``Is it normal that $M_I(RC)$ follows 
the local prescription and $(V-I)_0$ does not?''.

In summary, with currently available photometry, it is possible to obtain
the consistent RR Lyrae and clump giant distance scales that differ
by as much as 0.2 magnitudes. 
Furthermore, even the presented distance scales may require some additional
shift due to possible adjustments in $\alpha$, $M_I^{\rm BW}(RC)$, and
zero-points of adopted photometry.
It is clear that further investigations of population dependence of $M_I(RC)$, 
the Galactic bulge colors and the zero points of the LMC photometry are needed
to better constrain the local distance scale.

\acknowledgments

I would like to thank Andrew Gould for his valuable comments.
I am grateful to the referee whose suggestions improved the presentation
of the paper.
This work was performed under the auspices of the U.S. Department of Energy 
by University of California Lawrence Livermore National Laboratory under 
contract No. W-7405-Eng-48.

\clearpage

\end{document}